\documentclass[journal]{IEEEtran}

\ifCLASSINFOpdf
\else
   \usepackage[dvips]{graphicx}
\fi
\usepackage{url}

\usepackage{graphicx}
\usepackage{amsmath,hyperref}
\usepackage{multirow}
\usepackage{booktabs}
\usepackage{amssymb}  

\begin{document}

\title{Listening for "You": Enhancing Speech Image Retrieval via Target Speaker Extraction}

\author{Wenhao Yang, \IEEEmembership{Student Member, IEEE}, Jianguo Wei, \IEEEmembership{Member, IEEE}, \\
Wenhuan Lu\dag, \IEEEmembership{Member, IEEE}, Xinyue Song, Xianghu Yue, \IEEEmembership{Member, IEEE}
\thanks{This work was supported by National Key R\&D Program of China (No.2023YFB2603902) and the Major Science\&Technology Specific Project of XINING(No.2024-Z-7).}
\thanks{Wenhao Yang, Jianguo Wei, Wenhuan Lu, Xinyue Song, Xianghu Yue are with the College of
Intelligence and Computing, Tianjin University, Tianjin 300350, China (e-mail:
\{yangwenhao, jianguo, wenhuan\}@tju.edu.cn). }
\thanks{\dag Corresponding author: Wenhuan Lu.}}

\maketitle

\begin{abstract}

Image retrieval using spoken language cues has emerged as a promising direction in multimodal perception, yet leveraging speech in multi-speaker scenarios remains challenging. We propose a novel Target Speaker Speech-Image Retrieval task and a framework that learns the relationship between images and multi-speaker speech signals in the presence of a target speaker. Our method integrates pre-trained self-supervised audio encoders with vision models via target speaker-aware contrastive learning, conditioned on a Target Speaker Extraction and Retrieval (TSRE) module. This enables the system to extract spoken commands from the target speaker and align them with corresponding images. Experiments on SpokenCOCO2Mix and SpokenCOCO3Mix show that TSRE significantly outperforms existing methods, achieving 36.3\% and 29.9\% Recall@1 in 2- and 3-speaker scenarios, respectively—substantial improvements over single-speaker baselines and state-of-the-art models. Our approach demonstrates potential for real-world deployment in assistive robotics and multimodal interaction systems.

\end{abstract}


\IEEEpeerreviewmaketitle

\section{Introduction}

\IEEEPARstart{H}{uman} communication in real-world environments often involves multiple speakers, where listeners naturally focus on a target speaker while filtering out others—a phenomenon known as the \textit{cocktail party} problem \cite{6788551}. In human-computer interaction, particularly in multimodal retrieval tasks such as linking speech to images, this ability to isolate target speech is critical for both accuracy and security. Despite progress in speech-image retrieval, existing studies predominantly assume single-speaker inputs and have largely overlooked the challenge of mixed multi-speaker speech \cite{merkx2019language,sanabria2021talk,10022954}. This gap limits their applicability in realistic scenarios where overlapping speech is common, such as smart homes, meetings, or public spaces. 

Speech-image retrieval aims to align spoken utterances with corresponding images by learning cross-modal representations. Inspired by the success of CLIP \cite{Radford2021LearningTV} in text-image retrieval, recent works such as SpeechCLIP \cite{10022954} and AudioCLIP \cite{9747631} leverage pre-trained Self-Supervised Learning (SSL) models \cite{schneider19_interspeech,NEURIPS2020_92d1e1eb,hsu2021hubert} as speech / audio encoders. These models map speech into a shared semantic space with image representations—typically extracted from a frozen CLIP image encoder—via contrastive learning. During inference, speech and image embeddings are matched using similarity metrics such as cosine distance. However, these methods treat speech as a monolithic input, ignoring speaker identity and potential interference from non-target speakers.

Target speaker extraction (TSE) aims to isolate the speech of a specific target speaker from a mixture of overlapping voices by leveraging reference information (e.g., speaker embeddings or visual cues). Inspired by advancements in speech separation \cite{8707065} and speaker verification \cite{snyder2018x}, recent TSE methods such as VoiceFilter \cite{wang19h_interspeech}, TEA-PSE \cite{10446418} and WhisperTSE \cite{11015756} employ speaker-conditioned separation networks or generative models. These approaches typically employ the target speaker's enrollment speech to guide the separation process, where some methods leverage speaker embeddings derived from speaker verification models \cite{snyder2018x,wang23ha_interspeech}, whereas others rely on the input waveform. During inference, the model extracts the target speaker's features from the mixed signal using attention mechanisms or mask estimation \cite{11087624}, generating a clean speech output that focuses on the target speaker. Previous work has also explored TSE task in downstream applications such as multi-speaker ASR and speaker diarization \cite{10097139,10887683,10768974,10777072}.

In this letter, we extend speech-image retrieval to a more realistic and challenging setting: \textit{Target Speaker Speech-Image Retrieval} in multi-speaker environments. Given a speech mixture, our goal is to retrieve images corresponding only to a pre-enrolled target speaker, while suppressing interference from other speakers and background noise. We propose a framework that preserves the base model's general retrieval capability while enabling selective, speaker-conditioned retrieval. This work establishes a new direction for robust, speaker-aware multimodal systems that better reflect the complexities of natural auditory scenes.

We propose integrating target speaker extraction into the current single-speaker speech-image retrieval pipeline to enable target speaker awareness. In our framework, we employ a state-of-the-art speaker verification model to extract embeddings from enrollment utterances for target speaker extraction. The integration of target speaker information occurs offline, a limitation that warrants further investigation. Our main contributions can be summarized as follows:

\begin{figure*}[htb]
\centering
\includegraphics[width=0.9\linewidth]{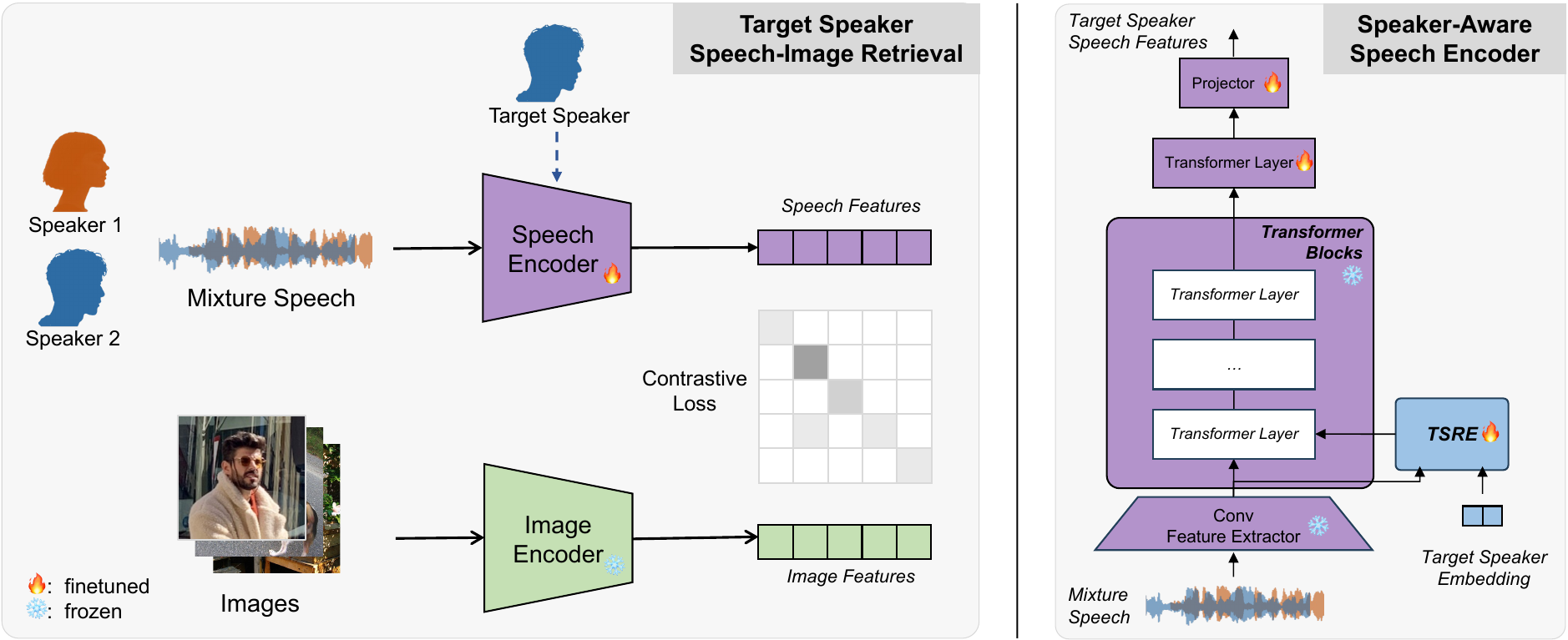}
\caption{The framework for Target Speaker Speech-Image Retrieval task. Left: the overall pipeline. Right: The Speaker-Aware SSL-based speech encoder.}
\label{fig:ts_clip}
\end{figure*}

\begin{enumerate}
    \item Introducing the novel Target Speaker Speech-Image Retrieval task and benchmark;
    \item Proposing the efficient, hot-swappable TSRE module for speaker-aware contrastive learning.
\end{enumerate}

\section{Methodology}
\label{sec:methd}

In this section, we introduce the Target Speaker Speech-Image Retrieval task and present a framework for it, as shown in Figure~\ref{fig:ts_clip}. We first describe the vanilla speech-image retrieval task. We then extend it to multi-speaker scenarios and introduce our proposed framework with a Target Speaker Retrieval Extractor module.

Following the CLIP framework, cross-modal retrieval can be achieved via contrastive learning, where image and speech inputs are projected into a shared semantic space. Their similarity is then measured using metrics such as cosine distance. The training objective for sample $m$ is defined as:
\begin{equation}
\label{eq4}
\begin{split}
\mathbf{e}_i &= \mathcal{E}_i(\boldsymbol{x}_i),  \mathbf{e}_s = \mathcal{E}_s(\boldsymbol{x}_s), \\
\mathcal{L}_{\text{i} \to \text{s}}^{(m)} &= -\log \frac{\exp\left(\sim({\mathbf{e}}_{i_m}, {\mathbf{e}}_{s_m}) / \tau\right)}{\sum_{n=1}^N \exp\left(\sim({\mathbf{e}}_{i_m}, {\mathbf{e}}_{s_n}) / \tau\right)} \\
\mathcal{L}_{\text{s} \to \text{i}}^{(m)} &= -\log \frac{\exp\left(\sim({\mathbf{e}}_{s_m}, {\mathbf{e}}_{i_m}) / \tau\right)}{\sum_{n=1}^N \exp\left(\sim({\mathbf{e}}_{s_m}, {\mathbf{e}}_{i_n}) / \tau\right)}
\end{split}
\end{equation}
where $\mathcal{E}_i$ and $\mathcal{E}_s$ denote the image and speech encoders, $\boldsymbol{x}_i$ and $\boldsymbol{x}_s$ are the image and speech inputs, $\mathbf{e}_i$ and $\mathbf{e}_s$ are fixed-dimensional embeddings, and $\tau$ is the temperature parameter.

In CLIP-based speech-image retrieval, the visual encoder is frozen during training. Prior works \cite{10022954,10625827} employ HuBERT \cite{hsu2021hubert} as the speech encoder.

\subsection{Target Speaker Speech-Image Retrieval}
\label{sec:format}

We formalize target speaker speech-image retrieval in multi-speaker scenarios using the following notation. The mixture speech input contains $K \geq 1$ overlapping speakers:
\begin{equation}
\mathbf{x}^K \in \mathbb{R}^T
\end{equation}
where $\mathbf{x}^K$ is a time-series audio signal. Multi-speaker speech-image retrieval aims to extract the target speaker's representation given a conditional identity $p \in \{1, 2, \dots, K\}$:
\begin{equation}
\mathbf{e}_s | \mathbf{u}^p  = \mathcal{E}_s'(\mathbf{x}^K, \mathbf{u}^p) 
\end{equation}
where $\mathbf{u}^p$ is pre-enrolled speaker identity information—such as a speech segment or fixed-dimensional embedding (computed via a speaker verification model)—and $\mathcal{E}_s'$ is a Speaker-Aware Speech Encoder as shown in Figure~\ref{fig:ts_clip}.

The target speaker speech-to-image contrastive loss for sample $m$ is:
\begin{equation}
\begin{split}
\mathcal{L}_{\text{i} \to \text{s}}^{(m)} = -\log \frac{\exp\left(\sim({\mathbf{e}}_{i_m}, {\mathbf{e}}_{s_m} | \mathbf{u}^p) / \tau\right)}{\sum_{q=1}^K \sum_{n=1}^N \exp\left(\sim({\mathbf{e}}_{i_m}, {\mathbf{e}}_{s_n} | \mathbf{u}^q) / \tau\right)} \\
\mathcal{L}_{\text{s} \to \text{i}}^{(m)} = -\log \frac{\exp\left(\sim({{\mathbf{e}}_{s_m} | \mathbf{u}^p, \mathbf{e}}_{i_m}) / \tau\right)}{\sum_{q=1}^K \sum_{n=1}^N \exp\left(\sim({\mathbf{e}}_{s_m} | \mathbf{u}^q,{\mathbf{e}}_{i_n}) / \tau\right)}
\end{split}
\end{equation}

The total training objective is the average of bidirectional losses over the batch:
\begin{equation}
\mathcal{L}_{\text{total}} = \frac{1}{2M} \sum_{m=1}^M \left( \mathcal{L}_{\text{i} \to \text{s}}^{(m)} + \mathcal{L}_{\text{s} \to \text{i}}^{(m)} \right)
\end{equation}
where $M$ is the batch size.

\subsection{Target Speaker Retrieval Extractor}
\label{sec:tsre}

\begin{figure}[htb]
\centering
\includegraphics[width=0.8\linewidth]{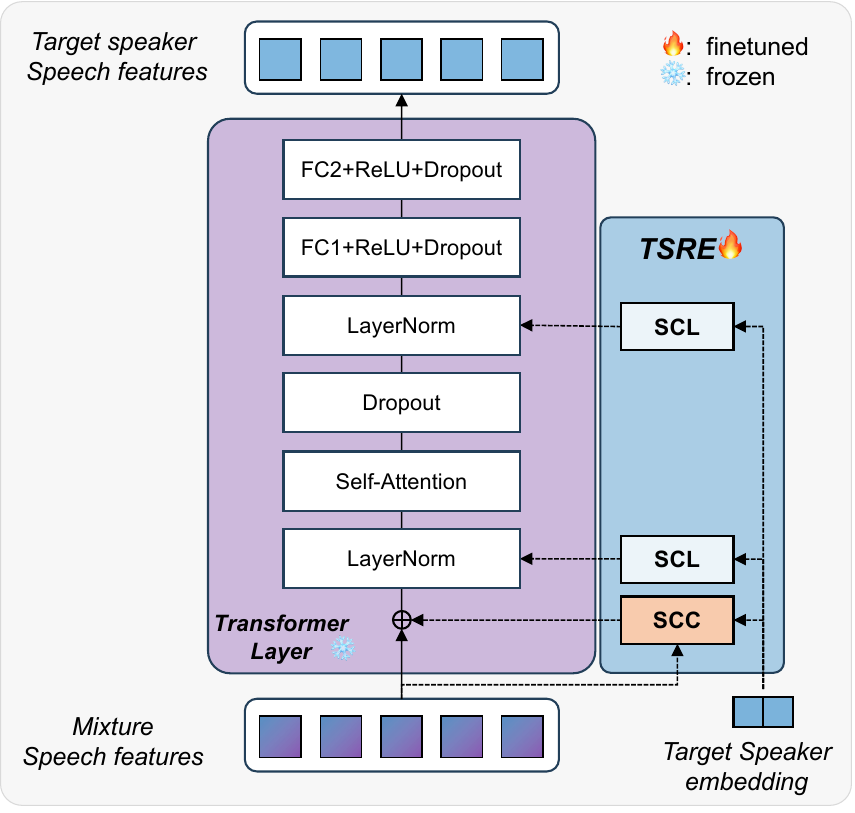}
\caption{Proposed Target Speaker Retrieval Extractor module for SSL models.}
\label{fig:SEM}
\end{figure}

Inspired by prior work on target speaker extraction in SSL frameworks, we propose a Target Speaker Retrieval Extractor (TSRE) module integrated into SSL models. The TSRE module consists of two types of components: Speaker-Conditional LayerNorm (SCL) and Speaker-Conditional Convolution (SCC), designed to capture long-term speaker normalization statistics and short-term speaker-specific details, respectively, as shown in Figure~\ref{fig:SEM}.

Speaker-Conditional LayerNorm (SCL), inspired by the CLN method~\cite{10097139}, employs a Feature-wise Linear Modulation (FiLM) mechanism~\cite{10.5555/3504035.3504518} to condition Transformer layer LayerNorm operations with speaker embeddings. Target speaker information modulates the normalization statistics of feature representations. Layer normalization is computed as:
\begin{equation}
\label{eq4}
\hat{\mathbf{h}} = \frac{\mathbf{h} - \mu}{\sigma} \cdot \gamma + \beta
\end{equation}
where $\mu$ and $\sigma$ are the mean and standard deviation of input $\mathbf{h}$, and $\gamma$, $\beta$ are learnable affine parameters. The FiLM module replaces $\gamma$ with a speaker-dependent scaling factor $\gamma'$:
\begin{equation}
\gamma' = w(\mathbf{u}) \cdot \gamma + b(\mathbf{u})
\end{equation}
where $\mathbf{u}$ is the speaker embedding.

To further enhance target speaker extraction, we introduce a novel Speaker-Conditional Convolution (SCC) module before SCL. This module applies a convolution operation to extract short-term features of the target speaker. The 1D convolution kernel weights are linearly modulated by the speaker embedding, similar to SCL. The hidden states are then updated via grouped convolution:
\begin{equation}
\hat{\mathbf{h}} = \mathbf{h} + s \cdot \text{Conv1D}(\mathbf{h}, w_c + \text{FC}(\mathbf{u}))
\end{equation}
where $w_c$ is the base convolution kernel, and $s$ is a learnable scaling factor initialized to zero.

Introducing the SCC module into Transformer blocks adds trainable parameters that depend on the dimension of hidden state features. However, the dimension of speech hidden states may significantly exceed that of speaker embeddings. To mitigate parameter growth and maintain architectural effectiveness, we propose inserting downsampling and upsampling pointwise and grouped convolutional layers before and after the SCC module, respectively. We denote this revised structure as the Speaker-Conditional Bottleneck Convolution (SCC-B) module:

\begin{equation}
\begin{split}
\bar{\mathbf{h}} &= \text{DownConv1D}(\mathbf{h}) \\
\tilde{\mathbf{h}} &= \bar{\mathbf{h}} + s \cdot \text{Conv1D}(\bar{\mathbf{h}}, w_c + \text{FC}(\mathbf{u})) \\
\hat{\mathbf{h}} &= \mathbf{h} + \text{UpConv1D}(\tilde{\mathbf{h}})
\end{split}
\end{equation}

\section{Experiments}
\label{sec:exp}

\subsection{Settings}
\label{sec:intro}

\noindent \textbf{Dataset} We use Flickr8k~\cite{harwath2015deep} and SpokenCOCO~\cite{hsu2021text} as training and evaluation datasets. Following target speech separation tasks, we synthesize a multi-speaker mixture corpus using SpokenCOCO and open-source code LibriMix~\cite{Cosentino2020LibriMixAO}. For speaker enrollment, we select 6-second utterances from each speaker in all Karpathy splits~\cite{7534740}. The pretrained speaker verification model, ECAPA-TDNN~\cite{desplanques2020ecapa}, is used to extract 256-dimensional speaker embeddings for the enrollment utterances. In our setup, mixtures are formed using only clean speech, without added noise. We create two mixture variants: a 2-speaker version (\textit{SpokenCOCO2mix}) and a 3-speaker version (\textit{SpokenCOCO3mix}). Details are provided in our code\footnote{https://github.com/Wenhao-Yang/TS-SpeechCLIP} and Table~\ref{tab:dataset}.

\begin{table}[h]
\centering
\small
\setlength{\tabcolsep}{4pt} 
\caption{The training subsets of multi-modal datasets for the Speech Image Retrieval task.}
\label{tab:dataset}
\begin{tabular}{lcccc}
\toprule
\multicolumn{1}{c|}{\textbf{Datasets}} & \textbf{\#Image} & \textbf{\#Utt} & \textbf{\#Spk/Utt} & \textbf{Hours}  \\
\midrule
Flick 8k          & 6,000   & 30,000  & 1 & 34.4  \\
SpokenCOCO        & 113,287 & 567,171 & 1 & 684.0 \\
\midrule
SpokenCOCO2Mix    & 57,830  & 254,200 & 2 & 368.3 \\
SpokenCOCO3Mix    & 57,830  & 254,200 & 3 & 396.8 \\
\bottomrule
\end{tabular}
\end{table}

\noindent \textbf{Models} We implement speech-image retrieval models based on SpeechCLIP~\footnote{https://github.com/atosystem/SpeechCLIP} code, extending our prior work YOSS~\cite{10889085} and leveraging fairseq~\footnote{https://github.com/facebookresearch/fairseq}. This framework employs an alignment loss between speech and text embeddings without introducing additional trainable parameters. We evaluate two baselines using HuBERT and WavLM~\cite{9814838}, both in their \textit{LARGE} variants with 314M parameters. Each SSL model is followed by a learnable weight-sum layer and one transformer layer, adding 13.4M trainable parameters. For the TSRE module adapted to the \textit{LARGE} model with 1024-dimensional features, the SCL and SCC components contribute approximately 1.05M and 1.59M parameters, respectively. The SCC-B module features a 512-dimensional hidden space, half that of SSL models (1024). Two variants exist: SCC-B5 (kernel size 5) and SCC-B3 (kernel size 3). Our primary result employs the SCC-B3 variant.

We use the Adam optimizer with an initial learning rate of 2e-5, a batch size of 256, and a weight decay of 1e-8. Single speaker retrieval models are trained for 50,000 steps, and multi-speaker retrieval models are initialized by single speaker retrieval models and TSRE modules are finetuned for 32,000 steps. The checkpoint with the best validation performance is selected for testing. 

\begin{table*}[h]
\small
\centering
\caption{Target Speaker Speech-Image Retrieval on SpokenCOCO2Mix and SpokenCOCO3Mix test sets. \textbf{bold}: the best Recall.}
\label{tab:spokenmix}
\begin{tabular}{l l ccc ccc | ccc ccc}
\toprule
\multirow{4}{*}{\textbf{Encoder}} & \multirow{4}{*}{\textbf{Method}} & 
\multicolumn{6}{c|}{\textbf{\textit{SpokenCOCO2Mix}}} & \multicolumn{6}{c}{\textbf{\textit{SpokenCOCO3Mix}}} \\
\cmidrule(lr){3-8} \cmidrule(l){9-14}
& & \multicolumn{3}{c}{Speech$\rightarrow$Images} & \multicolumn{3}{c|}{Image$\rightarrow$Speech} &
 \multicolumn{3}{c}{Speech$\rightarrow$Images} & \multicolumn{3}{c}{Image$\rightarrow$Speech} \\
\cmidrule(lr){3-8} \cmidrule(l){9-14}
& & R@1 & R@5 & R@10 & R@1 & R@5 & R@10 & R@1 & R@5 & R@10 & R@1 & R@5 & R@10 \\
\midrule
\multirow{3}{*}{HuBERT} & Base$^*$       &  5.8 & 14.1 & 19.1 & 11.3 & 33.6 & 43.5 &  
                    1.7 &  4.7 &  6.9 &  2.9 & 13.2 & 19.6 \\
& CLN~\cite{10097139} & 15.8 & 33.9 & 43.1 & 35.0 & 62.3 & 72.8 &  
                    5.1 & 13.1 & 18.1 & 19.4 & 40.0 & 50.7 \\
& \textbf{TSRE (Ours)}       & \textbf{28.1} & \textbf{54.7} & \textbf{66.0} & \textbf{45.4} & \textbf{72.8} & \textbf{83.1} &  
                   \textbf{15.2} & \textbf{33.4} & \textbf{42.8} & \textbf{35.0} & \textbf{62.4} & \textbf{73.7} \\
\midrule
\multirow{3}{*}{WavLM} & Base$^*$        & 12.6 & 25.3 & 31.4 & 19.5 & 53.8 & 65.1 &  
                    4.8 & 11.1 & 14.7 &  8.1 & 30.6 & 41.2 \\
& CLN~\cite{10097139} & 33.3 & 61.4 & 72.3 & 53.1 & 77.0 & 85.2 &  
                   21.2 & 42.5 & 52.3 & 43.7 & 69.9 & 79.9 \\
& \textbf{TSRE (Ours)}       & \textbf{36.3} & \textbf{65.1} & \textbf{76.1} & \textbf{55.3} & \textbf{77.5} & \textbf{86.3} &  
                   \textbf{29.0} & \textbf{55.3} & \textbf{66.3} & \textbf{49.8} & \textbf{73.4} & \textbf{83.0} \\
\bottomrule
\end{tabular}
\end{table*}

\begin{table}[h]
\small
\centering
\setlength{\tabcolsep}{2.75pt}  
\caption{Single Speaker Speech-Image Retrieval on the Flickr 8K and SpokenCOCO validation test set. \textbf{bold}: the best Recall.}
\begin{tabular}{lcccccc}
\toprule
\multicolumn{1}{c|}{\multirow{2}{*}{\textbf{Method}}} & \multicolumn{3}{c}{Speech$\rightarrow$Images} & \multicolumn{3}{c}{Image$\rightarrow$ Speech} \\
\multicolumn{1}{c|}{} & \multicolumn{1}{c}{R@1} & \multicolumn{1}{c}{R@5} & \multicolumn{1}{c}{R@10} & \multicolumn{1}{c}{R@1} & \multicolumn{1}{c}{R@5} & \multicolumn{1}{c}{R@10}\\
\midrule
& \multicolumn{6}{c}{\textit{\textbf{Flickr8k}}}  \\
\midrule
MILAN \cite{sanabria2021talk}               & 33.2 & 62.7 & 73.9 &  49.6 & 79.2 & 87.5  \\
SpeechCLIP \cite{10022954}                  & 39.1 & 72.0 & 83.0 &  54.5 & 84.5 & 93.2 \\
SpeechCLIP+ \cite{10625827}                 & 41.7 & 73.7 & 84.1 &  54.2 & 86.8 & 94.2 \\
CMD-SpeechCLIP \cite{zhou24c_interspeech}   & 40.7 & 75.1 & 85.8 &  56.8 & 86.2 & 94.2 \\
HuBERT$^*$ (Ours)                 & 51.3 & \textbf{82.1} & \textbf{90.2} & 69.5 & 93.2 & 96.8  \\
WavLM$^*$  (Ours)                 & \textbf{52.1} & 81.8 & 90.1 & \textbf{71.2} & \textbf{93.7} & \textbf{97.6}  \\
\midrule
& \multicolumn{6}{c}{\textit{\textbf{SpokenCOCO}}}  \\
\midrule
SpeechCLIP \cite{10022954}                 & 35.8 & 66.5 & 78.0 & 50.6 & 80.9 & 89.1 \\
Seg.SpeechCLIP \cite{bhati23_interspeech}  & 28.2 & 55.3 & 67.5 & 28.5 & 56.1 & 68.9  \\
SpeechCLIP+ \cite{10625827}                & 36.5 & 66.3 & 77.9 & 51.0 & 80.0 & 88.5 \\
CMD-SpeechCLIP \cite{zhou24c_interspeech}  & 37.5 & 67.3 & \textbf{78.6} & 52.3 & 81.4 & 89.7  \\
HuBERT$^*$ (Ours)  &  37.8 & 66.9 & 78.0 & 55.1 & 82.7 & 90.0  \\
WavLM$^*$  (Ours)  & \textbf{38.0} & \textbf{67.7} & 78.5 & \textbf{58.5} & \textbf{83.5} & \textbf{90.9} \\
\bottomrule
\end{tabular}
\label{tab:baseline}
\end{table}

\subsection{Results}
\label{sec:result}
For the multi-speaker speech-image retrieval task, we employ target speaker-conditioned speech embeddings for retrieval, where the image described by the target speaker serves as the target image. This process follows the standard single-speaker retrieval paradigm. When applying single-speaker retrieval models (Base$^*$) to multi-speaker scenarios, we directly utilize speech embeddings extracted from mixture signals as target speaker embeddings for retrieval. We report Recall@K (K=1,5,10).

\noindent \textbf{Target Speaker Speech-Image Retrieval}
We evaluate target speaker speech-image retrieval on SpokenCOCO2Mix and SpokenCOCO3Mix in Table~\ref{tab:spokenmix}. The CLN method~\cite{10097139} is equivalent to the SCL method in Section~\ref{sec:tsre}. Our proposed TSRE module, combined with target speaker contrastive learning, significantly improves retrieval performance in multi-speaker scenarios. Single-speaker models suffer significant degradation: for instance, the WavLM baseline's Recall@1 drops from 38.0\% (single-speaker) to 12.6\% (2-speaker) and 4.8\% (3-speaker) in speech-to-image retrieval. With TSRE, Recall@1 improves to 36.3\% and 29.9\%, respectively. TSRE achieves 3.0\% and 7.8\% higher Recall@1 than CLN, with gains of 12.3\% and 10.1\% in HuBERT-based models. Multi-speaker retrieval remains challenging, with performance degrading by 0--10\% on average as speaker count increases.

Moreover, WavLM consistently outperforms HuBERT, likely because it was pretrained with tasks related to mixture speech separation, making it better suited for multi-speaker scenarios. The performance gap between WavLM-based and HuBERT-based models becomes particularly pronounced in the 3-speaker setting.

\noindent \textbf{Single Speaker Speech-Image Retrieval}
We report single-speaker speech-image retrieval baseline results on Flickr8k and SpokenCOCO, as shown in Table~\ref{tab:baseline}. Our models achieve acceptable performance on speech-image retrieval benchmarks. 

\begin{table}[h]
\centering
\small
\caption{Ablation studies of TSRE components on SpokenCOCO2Mix. \textbf{bold}: the best Recall, \underline{underlined}: the second best Recall.}
\label{tab:ablation}
\setlength{\tabcolsep}{3.75pt} 
\begin{tabular}{lccccccc}
\toprule
\multicolumn{1}{c}{\multirow{2}{*}{\textbf{Module}}} &  \multicolumn{1}{c}{\multirow{2}{*}{\#Param(M)}} & \multicolumn{3}{c}{Speech$\rightarrow$Image} & \multicolumn{3}{c}{Image$\rightarrow$Speech} \\
 &  & \multicolumn{1}{c}{R@1} & \multicolumn{1}{c}{R@5} & \multicolumn{1}{c}{R@10} & \multicolumn{1}{c}{R@1} & \multicolumn{1}{c}{R@5} & \multicolumn{1}{c}{R@10}\\
\midrule
-             & -                  & 12.6 & 25.3 & 31.4 & 19.5 & 53.8 & 65.1 \\
SCL           & 1.05           & 33.3 & 61.4 & 72.3 & 53.1 & 77.0 & 85.2 \\
SCC           & 3.16    & 34.2 & 62.6 & 73.5 & 53.5 & 77.0 & 85.9 \\
SCC-B5        & 2.11    & \underline{35.6} & \underline{64.5} & \underline{75.4} & \textbf{55.3} & \textbf{78.2} & \textbf{86.9} \\
SCC-B3        & 1.59    & \textbf{36.1} & \textbf{65.1} & \textbf{76.0} & \underline{55.1} & \underline{78.1} & \underline{86.5} \\
\bottomrule
\end{tabular}
\end{table}

\noindent \textbf{Ablation Studies} 
To validate the proposed TSRE module, we conduct ablation studies on its submodules, as shown in Table~\ref{tab:ablation}. Both SCL and SCC variants contribute to target speaker speech-image retrieval, with SCC variants achieving better performance than SCL. Among the SCC variants, SCC-B3—matching SCL in parameter count—yields improved performance on multi-speaker retrieval. Their combination achieves the best results on SpokenCOCO2Mix. SCL contains approximately 1.05M trainable parameters and SCC-B3 1.59M, together constituting less than 1\% of the parameters in the speech SSL-based retrieval model (317.4M). Thus, the proposed TSRE module is efficient and hot-swappable.

\section{Conclusion}
\label{sec:conclu}
In this letter, we introduce the novel task of Target Speaker Speech-Image Retrieval: identifying images associated with a specific speaker’s speech in mixed multi-speaker environments. We propose a contrastive learning framework based on a speech SSL model, enhanced with a flexible Target Speaker Extraction and Retrieval (TSRE) module that enables speaker-aware retrieval within existing architectures. Using publicly available resources, we construct and evaluate datasets for this task, demonstrating the effectiveness of the proposed approach. Our results suggest that focusing on the target speaker’s speech in multi-speaker settings is promising for human-computer interaction applications—such as robotic interaction, object detection, and retrieval—and may enhance both safety and accuracy in real-world scenarios. A limitation is the offline use of speaker embeddings; future work will explore end-to-end online speaker conditioning.

{\small
\bibliographystyle{IEEEbib}
\bibliography{strings,refs}
}

\end{document}